\documentclass[graybox]{svmult}

\usepackage{type1cm}        
\usepackage{makeidx}         
\usepackage{graphicx}        
\usepackage{multicol}        
\usepackage[bottom]{footmisc}

\usepackage{xcolor}
\usepackage{newtxtext}       %
\usepackage[varvw]{newtxmath}       

\makeindex             

\begin{document}

\title*{Olfactory search}
\author{Antonio Celani and Emanuele Panizon}
\institute{Antonio Celani \at The Abdus Salam International Center for Theoretical Physics ICTP, Str. Costiera, 11, 34151 Trieste, Italy. \email{celani@ictp.it}
\and Emanuele Panizon \at The Abdus Salam International Center for Theoretical Physics  ICTP, Str. Costiera, 11, 34151 Trieste, Italy. \email{epanizo1@ictp.it}}

\maketitle
\abstract{
{The task of olfactory search is ubiquitous in nature and in technology, from animals in the quest of food or of a mating partner, to robots searching for the source of hazardous fumes in a chemical plant. 
Here, we focus on the algorithmic approach to this task: we systematically review the different olfactory search strategies. Special emphasis is given to the formal description as a Partially Observable Markov Decision Processes, which allows the computation of optimal actions and helps clarifying the relationships between several effective heuristic search strategies.}}

\vspace*{1cm}
\textit{If you do not expect the unexpected you will not find it, for it is not to be reached by search or trail} (Heraclitus)

\section{Introduction}
\label{sec:intro}

Olfactory navigation and search are the processes by which organisms harness their sense of smell to navigate their environment, locate resources, and communicate with conspecifics. Olfaction, the sense of smell, serves as a cornerstone of survival and reproduction for a multitude of species across the animal kingdom, from microscopic bacteria to large mammals, transcending the boundaries of size and habitat~\cite{wyatt03}.

The olfactory sensory organs allow to perceive and interpret chemical cues, whether to detect the scent of prey, find a mate, identify food sources, or avoid danger where visual and auditory cues may fall short or remain unreliable~\cite{carde08}.

For some animals, like rodents, olfactory navigation is indispensable for exploring maze-like environments, remembering spatial layouts, and locating hidden rewards. For others, such as moths, it allows to track the scent of a potential mate across vast distances. Marine animals use their olfactory sense to migrate across oceans to their natal streams for spawning. These diverse examples highlight the broad spectrum of olfactory navigation applications in the natural world~\cite{Baker2018species}.

Several challenges must be faced by organisms that rely on olfaction for their navigation tasks. Odor cues are dispersed by turbulence in the air or water, diffused by the environment, and mixed with other scent sources, leading to ever-changing odor landscapes. Navigating this complexity requires an impressive array of adaptations, from the intricate structure of an insect's antennae to the advanced neural system of a canine's brain~\cite{reddy2022olfactory}.

In addition to natural organisms, the principles of olfactory navigation have far-reaching applications in the development of autonomous systems and robotics. Robots equipped with odor sensors can mimic the abilities of natural navigators, allowing them tracking the scent of chemical leaks in industrial settings or searching for survivors in disaster zones. Understanding the mechanisms and challenges of olfactory navigation in nature contributes significantly to the development of these technologies.
Vice versa, understanding the algorithmic principles of olfactory navigation can shed light on biological search processes.

 A comprehensive exploration of olfactory navigation and search would span biology, sensory physiology, behavioral ecology, neurobiology, fluid dynamics, and computational science. More humbly, in this chapter we  focus on the algorithmic view of olfactory search. 
After a brief review of the basic biological and physical facts that form the backdrop of our discussion, we will move quickly into the world of search strategies. A hopefully useful map to navigate this subject is given in Figure 1.

 \begin{figure}[!h]
\centering
	\includegraphics[width=\linewidth]{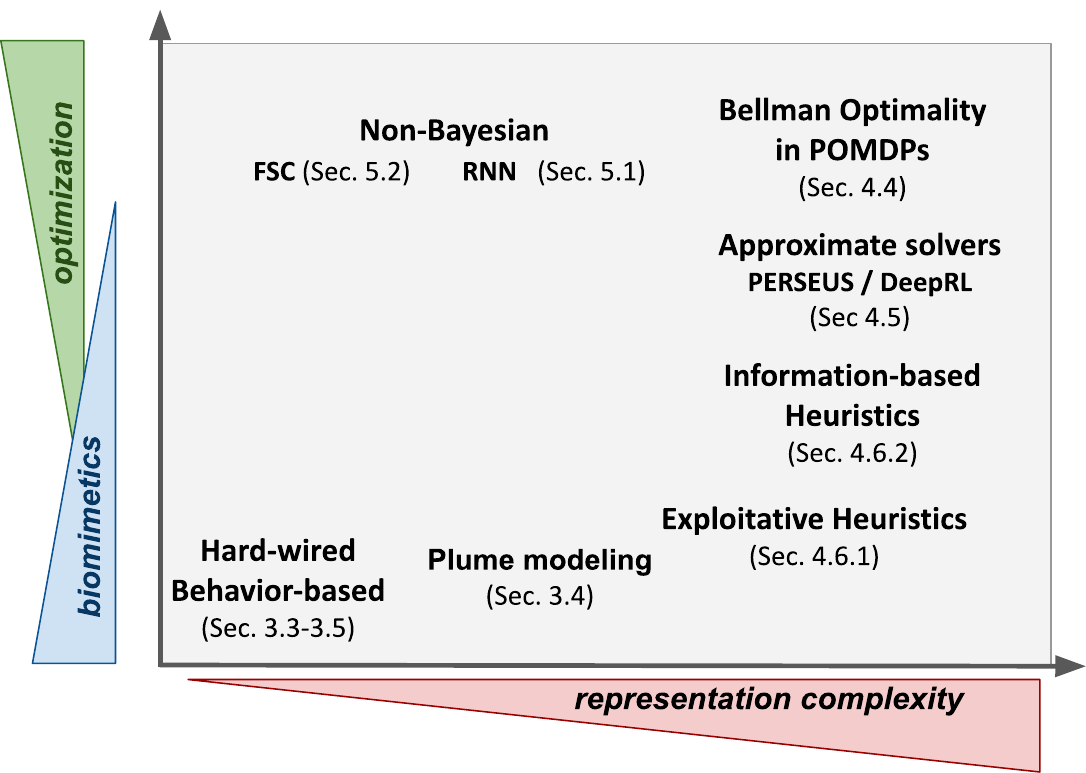}
	\caption{\label{fig:1} Concept map of the main algorithmic approaches to olfactory search discussed here and displayed on an abstract plane. On the $y$-axis search strategies are evaluated on how much they are grounded on principles of optimization or, conversely, rooted in biomimetics. The $x$-axis discriminates depending on the level of task-representation required by the agent: on the leftmost side, algorithms are purely reactive, while on the rightmost side full spatial maps and Bayesian inference is employed.}
\end{figure}

\section{The physics and biology of olfactory search}

Pheromones, specialized chemical compounds, serve as an astonishing example of long-range communication among various organisms. One of the most striking examples of this phenomenon can be observed in Lepidoptera, a vast order of insects that includes moths. These organisms display an exceptional ability to detect and respond to sex pheromones released by potential mates. 

In Lepidoptera, and many other species, the communication of sex pheromones is vital for successful reproduction. Most Lepidoptera can detect and be consistently drawn to calling females from considerable distances, sometimes spanning several hundred meters. What makes this achievement all the more impressive is that females release their pheromone messages into the turbulent atmospheric surface layer: males face the challenging task of deciphering the source's location from a signal that is not only attenuated by the environment but also garbled and mixed with various other olfactory stimuli. 

The pheromone communication system in moths, and other insects, operates under strong evolutionary pressure. Adult moths from certain families, such as Saturniidae and Bombycidae, have relatively short lifespans as adults. These moths primarily allocate their brief adult lives to reproduction, having stored up energy in the form of lipids during their larval stages. As a result, natural selection has shaped their olfactory systems to be exquisitely sensitive to pheromones. Just a few pheromone molecules coming into contact with a male moth's antenna are sufficient to alert the insect and trigger a change in its cardiac frequency. Even low concentrations of a few hundred molecules per cubic centimeter can elicit specific behavioral responses that precede flight.

Apart from the intensity of the pheromone signal, its quality and time-course play crucial roles. In terms of quality, the signal is typically a blend of two or more chemical compounds. Discrimination is achieved not by the components themselves but by different combinations and ratios within the mixture. Precision in discrimination is essential for finding a compatible mate.

The time-course of the pheromone signal is another crucial aspect. Turbulence in the environment has a profound impact, distorting the pheromone signal significantly. This leads to irregular and intermittent fluctuations in the concentration of pheromone at distances far from the source. The turbulent atmosphere causes the signal to feature alternating bursts of pheromone and clean-air periods, with a wide range of durations.

Characterizing the properties of odor detections in turbulent flows is a complex and fundamental problem in statistical fluid dynamics. Furthermore, the intermittency generated by the physics of turbulent transport is essential for eliciting the appropriate biological behavior. When insects are exposed to steady, uniform stimuli, they briefly move upwind, suspend their flight towards the source, and engage in crosswind casting, which is a typical response to the loss of olfactory cues. Males may temporarily resume upwind flight when the stimulus is increased incrementally or may engage in sustained upwind flight when exposed to repeated pulses. In these instances, the statistics of turbulence-airborne odor stimuli essentially constitute the message sent by females to male moths. This information governs their behavior, dictating how male moths respond and shaping their search strategies.

Understanding the statistics of odor detection during olfactory searches is essential for comprehending the neurobiological responses of insects. This intricate interplay between olfaction, turbulent fluid dynamics, and behavioral adaptations highlights the rich and multifaceted nature of pheromone communication in the natural world, offering insights into evolution, sensory biology, and the challenges faced by organisms in their quest for reproductive success.

The reader interested in the biological and physical mechanisms at work in olfactory search can find excellent introductions elsewhere~\cite{murlis1992odor,reddy2022olfactory,Baker2018species}. The characterization of the dynamical odor landscape in a turbulent environment  can be found in \cite{celani2014odor}.




\section{A day at the zoo}\label{sec:hardwired}

The olfactory search problem spans diverse research fields, from biology to robotics, where it is commonly known as Odor Source Localization (OSL). Its applications are broad, ranging from detecting hazardous gas leaks in power plants to gaining insights into animal behavior. Given the variety of perspectives applied to the olfactory search in recent decades, it is perhaps unsurprising that the methods and algorithms presented in scientific literature often lack a coherent structure.

In this section, we aim at summarizing -- without any pretense of an exhaustive review -- a wide array of methods proposed and analyzed to create efficient search strategies from a heuristic perspective. We will showcase examples of search strategies constructed with well-defined, hard-wired behaviors, which, although sometimes involving parameter inference, are not the outcome of an optimization process. 

In the subsequent sections, we will introduce a formal framework in which we will cast the olfactory search problem in the formal language of optimal decision-making in presence of uncertainty, i.e. the theory of Partially Observed Markov Decision Processes, detailed in \ref{sec:POMPDs}.
The latter will be the primary focus of this chapter. 

Readers who wish to take a deep dive in the robotic aspects of OSL can refer to \cite{kowadlo2008robot}, or to the more recent reviews in \cite{ristic2016study} and \cite{jing2021recent}.

\subsection{Zero information} 

An extreme case of search process is the one occurring under conditions of zero information, where the probability of encountering new odor signals is so low that it can be neglected.

One potential approach in such cases involves exhaustive searches, a concept that has also been explored in the context of naval warfare \cite{champagne2003search}. These strategies produce trajectories that attempt at a systematic coverage of all available space, either by gradually expanding in spiral-like patterns or by traversing parallel lines of fixed width. It is intriguing to compare the findings of this approach with the spiraling search patterns that emerge in the realm of ``optimal'' searches, as discussed in Section \ref{sec:linsearch}.

In a more biologically inspired context, it has been suggested that certain animals, like albatrosses, employ L{\'e}vy flights \cite{viswanathan1996levy} as a means to maximize the area they explore. However, the relative effectiveness of L{\'e}vy flights compared to potentially biased Brownian motion remains a topic of ongoing debate  \cite{Lomholt08,palyulin2014levy,levernier2020inverse}.
Connected to the notion of L{\'e}vy flights, an intermittent dual-mode strategy, combining fast-moving exploration and slow-moving detection, has been recognized as a widespread feature in target searches conducted under conditions of minimal information \cite{benichou2011intermittent}.

\subsection{Smooth odor landscapes}
Conversely, at the opposite end of the observational spectrum, we encounter what are often termed ``tactic'' methods of motion
\cite{jing2021recent}. Notably, these tactics do not necessitate an extended memory of previous observations, but rather rely on the continuous measurement of relevant field gradients, such as chemical concentration (chemotaxis), wind direction (anemotaxis), variations across the spatial sensor array (tropotaxis), or combinations thereof.

These tactic strategies appear in the natural world across a wide range of scales, from bacteria to ants and rodents. However, they typically falter in the presence of turbulence, where the environmental landscape is far from being smooth. A comparative review of tactic animal behaviors, contrasted with more complex strategies is given in \cite{Baker2018species}. 
Wadhwa and Berg~\cite{wadhwa2022bacterial} explore the mechanical aspects of motility in bacteria, shedding light on how chemotaxis is accomplished.

\subsection{Behavior-inspired search algorithms}
Expanding upon these tactical models of motion, researchers have introduced a diverse array of biomimetic algorithmic search patterns. These patterns often consist of straightforward, diagrammatic instructions, following the ``if-then-else'' format, which aim to replicate or imitate behaviors 
observed in nature. These algorithms rely on a series of pre-programmed behaviors, with the searcher switching between them based on immediate sensory inputs and/or predefined timing rules.

Some of these algorithms are designed to mimic purely reactive ``taxis''-like movements. For instance, there are Braitenberg-style tropotactic robots, equipped with two independent motors connected to sensors positioned at different locations. These robots utilize differences in signal intensity at the sensors to realign themselves toward or away from the signal source. In a more direct emulation of natural behavior, Holland and Melhuish~\cite{holland1996some} introduced an "E. coli algorithm" for a model run-and-tumble bacterium, while Sabelis and Schippers~\cite{sabelis1984variable} explored anemotactic strategies, {i.e., based on local wind direction,}  within odor plumes featuring variable wind directions. More recently, a tropotactic robot equipped with {displaced} gas sensors was developed~\cite{martinez2006biomimetic}, {exploiting the spatial fluctuations of the signal} .


Solutions based on hardwired behaviors have also been proposed to addressing the challenge of non-continuous signals, such as those presented by odor in turbulent environments. Typically, the initial efforts aimed at replicating the cast-and-surge behavior observed in moths, a behavior that had been increasingly scrutinized in controlled experimental settings \cite{kennedy1983zigzagging,ryohei1992self,kuenen1994strategies}.

The first generation of reactive methods, as proposed by \cite{kramer1997tentative}, involved alternations between upwind surges upon detecting odor and back-and-forth flights when no odor was detected. Building upon this fundamental concept, \cite{belanger1998biologically} introduced and compared several variants, including those with increased casting ranges. Interestingly, the introduction of varying casting ranges, dependent on the last detection, introduced the need for an internal clock and, consequently, a form of memory. This advancement expanded these algorithms beyond purely reactive approaches.

A more recent variation of these algorithms is the cast-and-surge algorithm developed by \cite{balkovsky2002olfactory}, where the alternating pattern of increasing casting range and surging motion demonstrated remarkable efficiency in reaching the source within (model) turbulent streams on a 2D square grid. \cite{shigaki2017time} introduced a behavioral model based on sequences of surges, zigzags, and loops, optimizing the timescale of transitions through experimentation with robotic setups.

In a different vein \cite{voges2014reactive} proposed the alternation of surging with spiraling and/or increased casting. They compared the performance of these methods with the more complex Infotaxis (refer to Section \ref{sec:infotaxis}) in experimental setups. Additional examples of diagram-based search in robotics can be found in \cite{russell2001tracking}, where a strategy that involved delayed turns and a discrete set of different motion modes was used to aid a robot in navigating maze-like environments, leveraging ``a combination of upwind search, detection, and knowledge of gross fluid dynamics.''

\subsection{Building a model of the odor plume -- and using it}

The algorithms discussed above primarily operate at a local level in their response, relying on pure reactivity or, at most, incorporating a rudimentary internal counter to modulate their search behavior. However, at a more advanced conceptual level, there are algorithms that involve some form of plume modeling. These ``cognitive'' methods are fundamentally grounded in a pre-constructed understanding of mechanism that produces  odor encounters. As signals accumulate over time, this model becomes increasingly defined, and the estimated source location converges to a smaller region. Therefore, the searcher faces two concurrent challenges: updating and refining the plume model and navigating within it.

It is important to remark that in this section, we will present methods that rely on plume modeling through techniques like parameter fitting or Bayesian inference. However, the navigation aspect is handled through hardwired behaviors. In the sequel (Section \ref{sec:Bellman} and Section \ref{sec:heuristics}), we will explore techniques where the navigation strategies are inextricably linked with the information-gathering process.

The initial step towards imbuing spatial ``awareness" in the agent involves constructing a real-time plume distribution based on parameter inference from the signals received by the agent as it moves through the environment. One of the early instances of such plume modeling can be found in \cite{ishida1997remote}, where the agent maintains a gas distribution model that is continuously fitted in real-time and subsequently used for odor source localization. In \cite{farrell2003chemical}, a combination of plume modeling through Hidden Markov States and behavior switching diagrams is applied for underwater Chemical Plume Tracing. Bayesian methods have also been employed for estimating chemical concentrations in water, as seen in \cite{pang2006chemical}, where inference was utilized to construct and update a source-likelihood map. This model incorporated inputs from concentration levels and fluid flow velocity, although it focused primarily on the ``mapping" aspect of the problem, neglecting the search strategy.

Furthermore, \cite{li2011odor} presents a robotic OSL algorithm based on particle filters. When odor signals are detected, the searcher engages in predefined exploratory behavior and employs the input signal to update the estimated source locations.

\subsection{Modeling search strategies from experimental data}

Recent advancements in real-time insect tracking technology allow to record and analyze experimental insect trajectories under various wind conditions. These trajectories have served as the foundation for developing behavioral models that elucidate the responses of insects to odor traces within turbulent flows. While these models may not explicitly represent search behavior, they hold considerable relevance for comparison, particularly when juxtaposed with heuristic or hardwired algorithms.

In \cite{alvarez2018elementary} experimental trajectories of walking Drosophila melanogaster were utilized to infer a behavioral model describing their responses to odor inputs within wind chambers. By incorporating adaptation and temporal filtering, the researchers proposed two-mode models where the history of the odor signal is transformed into ON/OFF activations. Various models were suggested, with the two modes corresponding to either an increase or decrease in ground speed accompanied by an opposing decrease or increase in turn rates. Another model variant involved different modulations dependent on the fly heading.

A subsequent work by \cite{matheson2022neural} introduced a ``neural circuit"-based model for navigation, incorporating inputs such as heading, wind direction, and odor, and simulating the fly response through layers of neural-inspired computations.

In \cite{demir2020walking} flies were observed as they responded to a smoke plume that attracted them. The flies exhibited a limited set of discernible actions, including ``stop," ``walk," and ``turn" (either upwind or downwind). The authors successfully quantified the statistics of these actions in conjunction with the local smoke concentration, employing a straightforward model based on transition rates between these actions.

\cite{kadakia2022odour} expanded upon the analysis of Drosophila trajectories, constructing a more effective model that incorporated directional cues related to odor motion.

\section{The optimality perspective of olfactory search}\label{sec:task}

The preceding section aimed at providing a non-exhaustive overview of the extensive efforts, spanning several decades, in the pursuit of efficient solutions to the olfactory search problem. What these proposals share is the reliance on well-defined sequences of behaviors, crafted through intuition, ingenuity, and common sense. While their performance arises from the precise combination of these features, none of these methods claim to be, nor are they, inherently ``optimal".

A distinct approach exists that frames olfactory search as a quest for an ``optimal" behavior within a rigorously defined formal framework, and this approach constitutes the primary focus of this section. The quest for {\it optimal} search strategies can be marked by a departure from human intuition. In optimal strategies, the selection of actions is inherently balanced between {\it exploration}, aimed at reducing the uncertainty of source location, and {\it exploitation}, directed at closing in on the likely source position.

Given the stochastic nature of signaling and the uncertainty regarding source location, olfactory search naturally falls within the domain of Partially Observable Markov Decision Processes (POMDPs), as detailed in Section \ref{sec:POMPDs}. Within the framework of POMDPs, we redefine the concept of spatial maps associated with ``cognitive" approaches using probabilistic density maps referred to as {\it beliefs} representing the ``perfect" encoding of past histories through Bayes' updates.

This section delves into methods that aim to (approximately) achieve optimality, as expounded in Section \ref{sec:Bellman}, while simultaneously addressing their strengths and limitations in the context of olfactory search. Intriguingly, the redefinition of the problem in terms of beliefs has served as a crucial starting point for the formulation and comprehension of several heuristic methods (Section \ref{sec:heuristics}). These heuristic methods, valued for their simplicity and practicality, continue to be some of the most widely applied techniques in olfactory search.

\subsection{The special case of Zero Information}\label{sec:linsearch}

Before describing the general framework of POMDPs, we discuss one of the extreme ends of the olfactory search problem. 
This is the case where the odor is so rarefied that no new olfactory signal becomes available to the agent during the search. The task is then to find a hidden source which is distributed according to a probability density function $b(s)$, typically peaked at the initial position of the search. 

In two or three dimensions, it is expected that the optimal strategy relies on exhaustive searches by non-self-crossing paths, such as outward-going spiraling. While it is known how heuristic strategies perform in this case, finding an (approximate) optimal solution even for 2D grid-worlds requires approximate methods (to be described in the next sections).

A special case, however, is the case of one-dimensional search, where this problem has been widely studied under the name {\it Linear Search Problem} (LSP)\cite{alpern2006theory}. On a line, any search strategy effectively amounts to a sequence of turning points $x_0 = 0, x_1, \dots, x_n,\dots $.  {The searcher starts in $0$, then moves straight to $x_1$, reverses direction and after crossing the origin reaches $x_2$, reverses again and so on and so forth until the source is found.} We can define the {expected} cost of a strategy as the space traveled to reach the source located in $s$. This strategy cost reads 

\begin{equation}
C = s \mathbf{1}(x_0 \leq s < x_1) + (|s| + 2 x_1) \mathbf{1}(x_2 \leq s < x_0) + (s + 2 x_1 + 2|x_2|) \mathbf{1}(x_1 \leq s < x_0) + \dots
\end{equation}

 \noindent where $\mathbf{1}(\cdot)$ is the indicator function, equal to one if the argument is true and zero otherwise, and we assumed that odd turning points are positive and even ones negative. If we define $B(x) = \int_{-\infty}^x b(s)\text{d}s$, it can be shown {by requiring $x_n$ to be a stationary point of the average cost, i.e., $\partial C/\partial x_n = 0$,} that the turning points which minimize the strategy cost must satisfy the following recursion for jump lengths

\begin{equation}
|x_{n+1} - x_n| = \frac{1-|B(x_n) - B(x_{n-1})|}{b(x_n)},
\end{equation}

Notably, for a few selected functional forms for $b_0(s)$, an analytical solution to the recursion can be obtained, which produces the optimal search strategy. If the starting distribution falls as a power law $b(s) \sim s^{-\alpha}$, for example, the resulting sequence of turns is geometric $|x_n| \sim \rho^n$, where $\rho$ satisfies the relation 
\begin{equation}
1+\rho =\frac{1}{\alpha}\left(1+\rho^\alpha\right),
\end{equation}  

For a general $b(s)$ the recursion cannot be solved in continuous space. On an infinite linear grid, however, it can always be solved up to any small approximation error~\cite{kao1997algorithms}. This is possible since one can always truncate the distribution on a finite segment, leading to a controllable error and then solve the bounded, finite-horizon task with standard methods of Dynamic Programming.

\subsection{Partially Observable Markov Decision Processes}\label{sec:pomdps}\label{sec:POMPDs}

The  olfactory search task falls in the wider framework of Partially Observable Markov Decision Processes (POMDPs). For a comprehensive discussion of POMDPs, we refer readers to \cite{kaelbling1998planning}. In this section, we will introduce the minimal elements required to establish a language to address algorithms within the field of olfactory search.

In POMDPs, the task is composed by a state space $S$, the set of actions $A$ and observations $O$, and a reward function $R:S\times A \rightarrow \mathbb{R}$. The agent at each time step selects an action $a \in A$ to move from its state $s \in S$  according to a transition probability $Pr(s'|s,a)$, which defines the physical environment. As it moves, it perceives an observation $o \in O$ and receives a reward $R(s,a)$ (discounted in time by a factor $\gamma$).
Since we are interested in minimizing the search time, a common reward function is to assign a positive reward of $1$ at the end of the task, when the source is found.

The task is then defined as an optimization problem, where the aim of the agent is to construct a \textit{policy}, i.e, the search strategy in our case, that maximizes the expected cumulative reward. In POMPDs the searcher does not have a complete knowledge of its state $s$ - which in the olfactory search case includes both the position of the agent and of the source. As a consequence, the policy in principle should map the whole history of past observations and actions $h_t = \{a_0, o_0, a_1, o_1,\dots,a_t,o_t\}$ 
to the probability of selecting an action at time $t+1$, i.e., $\pi(a|h_t)$. 
The quantity to maximize is the expected total reward $J$

\begin{equation}\label{eq:totrew}
J = \mathbb{E}[R_{tot}] =  \mathbb{E}\left[\sum_{t=0}^{\infty} \gamma^t R(s_t,a_t)\right] = \mathbb{E}[\gamma^{T-1}],
\end{equation}

\noindent where $T$ is the arrival time to the source.  
Explicitly constructing a policy on the past histories is almost always unfeasible, since histories grows exponentially in time. To solve this issue, several methods are available to construct tractable object which can in general be considered \textit{memories}, i.e., abstract representation of the past histories in a possibly reduced space. This can be obtained by exploiting the full knowledge of the observation model, as in Bayes' propagation of beliefs, but also in a model-free framework.

\subsection{Beliefs and cognitive spatial maps}

A formal solution of the above problem of encoding past histories of observations is through Bayes' propagation of belief. The agent constructs and evolves a \textit{belief} $b(s)$ which is a probability distribution over the state space $S${, here assumed to be discrete and finite}.
The belief lives in the (possibly very highly dimensional) simplex $b \in \Delta^{|S|-1}$, and is a sufficient statistic of the past history of observations and actions. For navigation tasks, and in particular for olfactory search, the belief is a probabilistic spatial map that encodes the probability that the target is at a given location. 

The policy $\pi(a|b)$ is a probability distribution over actions defined for any belief.

Beliefs are updated according to the Bayes' rule:
\begin{equation}\label{eq:belief_update}
b_{o,a}(s')  = \mathcal{T}^{o,a} (b) = \frac{Pr(o|s',a) \sum_{s} Pr(s' | s,a) b(s) }{\sum{s, s'} Pr(o|s',a) \sum_{s} Pr(s' | s,a) b(s)},
\end{equation}

Typically in the olfactory search literature the transitions - encoded in $Pr(s' | s,a)$ - are taken to be deterministic, expressing the approximation that the searcher has perfect control of its own movements. While this is not a necessary assumption, it allows to clear the stage from secondary issues other than the sensory/memory aspects.
Eq.~(\ref{eq:belief_update}) defines the operator of belief update $\mathcal{T}^{o,a}$.
Note that for the agent to construct the evolution of belief it is necessary to know the observation model $P(o|s', a)$. Moreover, the agent must have an initial belief $b_0(s)$, whose initialization will influence the performance and behavior of the searcher. 

A relevant quantity is the \textit{value} $V_\pi(b)$ of a policy, which is the expected reward that an agent with belief $b$ will obtain following a policy $\pi$. In our case  $V_\pi(b) =  \mathbb{E}_{a \sim \pi}[\gamma^{T-1} | b_0 = b ]$.

{ \color{black} \subsubsection{Models of plume}

We have now to construct a model for the observation $P(o|s,a)$. To do so, we rely on results on concentration fields in presence of a turbulent flow to produce a stationary probability distribution of odor encounters in any point in space. The advection-diffusion equation for the concentration field of  an odor source with emission rate $R$ and a turbulent flow with mean wind $V \hat{y}$ reads

\begin{equation}\label{eq:advdiff}
\partial_t c+V\partial_x c = D\nabla^2 c + R\delta(\mathbf{r})-c/\tau,
\end{equation}

\noindent where $c$ is the concentration field and $\tau$ the lifetime of the odor particle.
Eq.~\eqref{eq:advdiff} can be solved for its stationary solution in arbitrary dimensions 
In particular the solution in two dimensions is
\begin{equation}
c(\mathbf{r}) = \frac{R}{2\pi D}\text{e}^{\frac{-V(y-y_0)}{2 D}} K_0\left(\frac{|\mathbf{r}-\mathbf{r}_0|}{\lambda}\right),
\end{equation}

\noindent where $K_0$ is the {zero-th order} modified Bessel function and $\lambda = \sqrt{D\tau/(1+\frac{V^2 \tau}{4D})}$. From the mean concentration, one can obtain the average number of detections during a time $\Delta t$: $h(\mathbf{r})=c(\mathbf{r})D \Delta t 4 \pi l$, $l$ being the characteristic size of the searcher. In this case, therefore, we can extract the number of detections $o$ using a Poisson distribution with average number $h$: $P(o | s,a ) = \frac{h^o \exp(-h)}{o!}$. While this is an obvious over-simplification, this model is very useful in model-based techniques. 


The  model above, however, can only reproduce the mean rate of detection as a function of the position. Clearly, it will neglect higher space- and time-correlations. Unfortunately, while a full understanding of these properties remain a challenging and fundamental problem in statistical fluid dynamics~\cite{celani2014odor}, correlations are increasingly seen as a key ingredient of many animal responses to odor cues. For example it was shown that the \textit{Drosophila} flies respond with an upwind motion only to \textit{pulsed} carbon dioxide stimuli~\cite{zocchi2022co2}. 

Nonetheless, the performance of searchers optimized in a stochastic environment can be put to the test against realistic turbulent flow obtained via numerical simulations, or via imaging of experimental experiments, such as smoke puffs in wind tunnels~\cite{demir2020walking} or fluorescent dyes in water~\cite{duplat2010nonsequential,kree2013mixing} or in simple particle-based models for dynamical plumes in the presence of wind, as in \cite{singh2023emergent} or \cite{balkovsky2002olfactory}. 
}

\subsection{Optimality and Bellman equation}
\label{sec:Bellman}

We define $V^*$ as the  value function of the \textit{optimal} policy  $\pi^*$, which maximizes the expected total reward. $V^*$ satisfies the Bellman equation

\begin{equation}   \label{eq:Bell}
V^*(b) = \max_{a \in A}\left[ \sum_s b(s) R(s,a) + \gamma \sum_{o \in O} Pr(o|b,a) V^*(b_{o,a})\right]
\end{equation}

\noindent  where $Pr(o|b,a)$ is the probability of observing $o$, given the distribution $b$. If the optimal value $V^*$ is known, the optimal policy can be easily obtained by $\pi^*(b) = \arg \max_a \left[ \sum_s b(s) R(s,a) + \gamma \sum_{o \in O} Pr(o|b,a) V^*(b_{o,a})\right]$. 

It will be sometimes more convenient to use a related quantity, the optimal \textit{quality} function of the belief-action pair $Q^*(b,a)$. This function is related to the (optimal) value function by:

\begin{equation}
Q^*(b, a) = \sum_s b(s) R(s,a) + \gamma \sum_{o \in O} P(o|b,a)  V^*( b_{o,a} ).
\end{equation}

\noindent $Q^*(b,a)$ is the optimal discounted expected return after taking action $a$ and subsequently following the optimal policy.

When the observation corresponds to the source location, the POMDP actually reduces to a fully observable MDP. This case is just the specific instance where the belief is concentrated in a single point $s$, $b(s) = \delta(s)$. Since there is no need for further exploration  the optimal solution is in general trivial and the trajectory is given by the shortest path to the source. In the following we will refer to this case with $V_{\text{MDP}}(s) = V^*\left( b = \delta(s)\right)$ and the optimal action as $\Pi_{\text{MDP}}(s)$, following the convention in ~\cite{fernandez2006heuristic}.

\subsection{Approximate solvers}\label{sec:apprsol}

The Bellman optimality equation, Eq.~\eqref{eq:Bell} can in principle be solved by an iterative process, which is guaranteed to converge on the optimal value function. Introducing the Bellman operator $\mathcal{B}(\cdot)$, defined on the space of the value functions of beliefs,

\begin{equation}
\mathcal{B}(V) =  \max_{a \in A}\left[ \sum_s b(s) R(s,a) + \gamma \sum_{o \in O} Pr(o|b,a) V (b_{o,a})\right],
\end{equation}
 \noindent one has that Bellman operators are contractions, so that starting from any initial guess $V_0(b)$, one could iteratively construct $V_1(b) = \mathcal{B}(V_0)$, $V_2(b) = \mathcal{B}(V_1)$ and so on, that must converge to a fixed point. This \textit{value iteration} process is of practical use when the state space is discrete and relatively small. The high-dimensionality and the continuity of the belief space of a POMDP, however, makes the update step over all the belief space intractable even for the smallest state spaces $S$. In the following section we will describe two different approaches, where this ``dimensionality curse" is tackled either by  point-based methods in belief space, or by a functional approximation via Artificial Neural Networks (ANN). 

\subsubsection{DeepRL solver with Bellman Error minimizations}

The Bellman equation, Eq.~\eqref{eq:Bell}, holds for the optimal value function $V^*$. For any other function of the belief the l.h.r and r.h.s. of the equation will have a mismatch called Bellman Error. 
Recently \cite{loisy2022} proposed to find the optimal policy using value functions parametrized by a Neural Network with weights $w$, $V(b;w)$, and rephrasing the optimization task as finding the $w^*$ that minimize of the residual Bellman Error. This error reads

\begin{equation}
\mathcal{L}(w) = \mathbb{E}_{b}\left[ \max_{a \in A} \left( \sum_s b(s) R(s,a) + \gamma \sum_{o \in O} Pr(o|b,a) V(b_{o,a};w)\right) - V(b;w)\right]^2,
\end{equation}

In principle, the expectation could be taken over the whole space of beliefs, since the residual error of the true optimal value function $V^*(b)$ is zero everywhere. However, $V(b;w)$ is an approximation and the expectation in \cite{loisy2023deep} runs over the beliefs $b$ effectively visited by the searcher. The optimization of $w$ is done using  the functional $\mathcal{L}(w)$ as a ``loss function'' to be minimized through standard stochastic gradient descent, via back-propagation methods. 

This method was first proposed in \cite{loisy2022} and then further exploited in \cite{loisy2023deep} to solve the case of stochastic plumes in $1D$ and $2D$. The results showed that it converges to solutions outperforming other heuristic methods, and is competitive with other methods to reach optimized solutions~\cite{heinonen2023optimal}. 

\subsubsection{Approximate value iteration with Perseus}\label{sec:perseus}

The Perseus algorithm~\cite{spaan2005perseus} is based on \textit{point-based value iteration}, where the value function is parameterized by a finite number of vectors in the belief space. At each iteration step $n$, Perseus performs value iteration on a large sample of beliefs, using a finite set of vectors $\mathcal{A}^n$. For each belief it approximates the value function as a piecewise linear, convex function:

\begin{equation}
V(b) = \max_{\alpha \in \mathcal{A}^n} b \cdot \alpha,
\end{equation} 

\noindent where $\cdot$ indicates the dot product over all states $s$. The approximation is then plugged into the iteration - i.e., in the last term of Eq.~(\ref{eq:Bell}) - where it can be written in terms of \textit{propagation} vectors $g^i_{a,o} = \sum_{s'\in S}Pr(o|s',a)Pr(s'|s,a)\alpha_i(s')$. The vectors $g^i_{a,o}$ encode the evolution of the belief as a consequence of action $a$ and observation $o$ and need to be computed and stored only once, after constructing the belief set.
Each iteration can then be shown to consist in $\textit{backup}$ operations where the vectors $\alpha$ are updated,

\begin{equation}
\alpha_a' = R(s,a) + \gamma \sum_o \arg \max_{g^i_{a,o}} b \cdot g^i_{a,o}.
\end{equation}

Since each $\alpha$-vector is associated to an action, each belief is mapped into the action of its corresponding maximizing $\alpha$. The order of the backup operation matters, and it has been found that performing it in order of decreasing Bellman error offers a better convergence rate~\cite{heinonen2023optimal}.

The Perseus algorithm was applied to solve the olfactory search task of a odor plume in a turbulent flow~\cite{heinonen2023optimal}, where its efficiency is compared to several other heuristics, as discussed in the following sections. Rigolli \textit{et al.}~\cite{rigolli2022alternation} utilized the same algorithm to study the \textit{alternation} model of olfactory search. There, the search process involves two different modes of odor sampling, mimicking the rodent behavior that alternates between sniffing on the ground while moving, or stopping to sniff the air.
Since odor puffs travel differently close to the surface, where flow is laminar, than farther in the air the agent essentially can probe the environment through two different communication channels (in POMDP terms, two different functions $Pr_{x}(o|s,a)$, where $x \in \{\text{ground}, \text{air}\}$).

\subsubsection{Parameterizing the belief space}

A different approach to tackle the curse of dimensionality of the belief space was used by \cite{reddy2022sector} for the task of tracking a surface-bound odor trail. There, the posterior is propagated through standard Bayes updates, but it is parameterized by a mixture of $k$ Gaussian basis function. Using this reduced basis, the authors are able to use the SARSA Reinforcement Learning algorithm to learn the quality function $Q_\pi$, and successfully reproduce the characteristic crisscrossing patterns of trail searching.

\subsection{Bayesian heuristics}\label{sec:heuristics}

The complexity of solving the Bellman optimality equation for POMDPs has spurred efforts to produce effective algorithms based on spatial beliefs, without necessarily solving any optimization task. These methods are generally called ``heuristics" in the language of POMDPs (not to be confused with the usage of the same term in robotics).

Some of the following methods can also be recast in terms approximation of the value- or quality-functions $V(b)$, or the quality-function $Q(b,a)$, in terms of their equivalents for the case of full observability, i.e. $V_{\text{MDP}}(s)$ and $Q_{\text{MDP}}(a, s)$ and the relative optimal solution $\Pi_{\text{MDP}}(s)$.


\subsubsection{Simple heuristics}\label{sec:simpleheu}

A selection of the simpler heuristics can be found in ~\cite{fernandez2006heuristic}, and their performance in a virtual environment (and against more advanced methods) can be evaluated using the codes freely available in the OTTO package~\cite{loisy2022otto}.

The simplest method, called \textit{Most Likely State} (MLS)~\cite{cassandra1996acting}, prescribes that the action to be taken given a belief $b$ is simply to move in the direction where the probability to find the source is highest. In other terms $Q(b, a)$ {is replaced by} $ Q_{\text{MDP}}(\arg \max_s b(s),a )$.
A common variation relies on Thompson sampling, which relaxes the requirement of moving towards the most-likely state only. In Thompson sampling, the agent samples a position for the source with probability given by $b$, and then selects the optimal action for that location. This is equivalent to selecting actions randomly following the expected optimal actions $\pi_{\text{Thompson}}(a|b) =  \sum_s b(s) \mathbf{1}\left( a = \Pi_{\text{MDP}}(s)\right)$. 

Other methods include Action Voting - which is equivalent to selecting the most probable action as defined for the expected-Thompson algorithm - and QMDP, where the policy is constructed assigning each action to the corresponding average $Q_{\text{MDP}}$ value, weighted by the belief:  $\pi_{\text{QMDP}}(b) = \arg \max_a  \sum_s b(s) Q_{\text{MDP}}(s, a)$.

While relatively efficient at a very low computational cost, all these methods are essentially exploitative since they do not consider, neither explicitly nor implicitly, the benefit of performing actions to reduce uncertainty. Moreover, they are known to suffer from a sort of ``Buridan's ass syndrome'' when the belief landscape has multiple maxima and the agent is equidistant to many of them. Since observations are stochastic, the relative intensity of the maxima can fluctuate rapidly in time and the resulting back-and-forth motion can lead to severe performance loss. A possible solution to this is to permit time-persistent action selection~\cite{russo2018tutorial}, at the cost of introducing additional hyper-parameters to the algorithm.

\subsubsection{Information-seeking methods}\label{sec:infotaxis}

As mentioned, the previous methods do not correctly address the exploration part of the delicate exploration/exploitation balance required for optimality. In this section we will deal with information-seeking methods, which introduce an explicit push for exploration.

Several heuristics have been proposed which consider the (expected) effect of actions on the future beliefs. In particular, this can be achieved by rewarding the agent with the minimization of the expected entropy of the belief $b(s)$ at each step. While this method was first proposed as the ``uncertainty reducing'' mode in the Dual mode control~\cite{cassandra1996acting,fernandez2006heuristic}, it is most commonly referred  to by the name ``Infotaxis'' after its more recent application to the olfactory task in a turbulent medium~\cite{vergassola2007}.


The Infotaxis heuristic is based on taking actions that minimize the uncertainty on the source location. Such uncertainty is measured by the Shannon entropy of its distribution, i.e., of the belief, as $H(b) = - \sum_s b(s) \log_2 b(s)$. At each time, Infotaxis selects the action that maximizes the \textit{expected reduction of the belief's entropy} $G(b,a)$:

\begin{equation}
G(b,a) =  H(b) - \sum_o Pr(o|b,a) H(b_{o,a}).
\end{equation}

Contrary to the previous methods, Infotaxis is biased towards exploration against exploitation. In particular, it has no incentive to move once it has removed all uncertainty over its position and tends on err on side of curiosity-driven behavior. The Dual mode control~\cite{cassandra1996acting,fernandez2006heuristic} counteracts this deficiency by imposing a mode switch: above a certain threshold on the entropy, actions are selected to minimize uncertainty, whereas below it the behavior switches to any of the exploitative heuristics, such as Thompson sampling, QMDP or Action Voting.

A recent variant, called Space-Aware Infotaxis~\cite{loisy2022} addresses this issue differently, reformulating the quantity to minimize. The new objective is 

\begin{equation}
G_{SAI}(b,a) = - \sum_o Pr(o|b,a) \left( \text{e}^{H(b_{o,a})} + D(b_{o,a}) \right),
\end{equation}

\noindent where $D(b) = \sum_s b(s) ||x-x_a||$ {and the state $s=(x, x_a)$ contains the positions of the searcher $x$ and of the source $x_a$}. The second term measures the expected distance between the agent 
and the source 
, given $b$. Its minimization, i.e. the exploitation of the current belief $b$, counter-balance the information-gathering aspect of Infotaxis. Note that the exponentiation of the Shannon entropy, without the addition of the distance term, in principle does not modify the policy of Infotaxis since any ordering in the actions is preserved by it. However, it allows for a more direct interpretation, since it can be shown that $\text{e}^{H(b)}$ is strictly related to the minimum expected time needed to explore a distribution $b$ when arbitrary jumps are allowed. 



The uncertainty about the position of the source is not the only one that can be addressed.  In their original paper~\cite{cassandra1996acting} stipulate that there is a ``benign entropy'', where the uncertainty is between states which however share the same (MDP) optimal action. They therefore propose a Dual Mode algorithm based on the entropy of action, instead of state. Given $\phi(a) = \sum_s b(s) \mathbf{1}\left(a = \Pi_{\text{MDP}}(s)\right)$ to measure the relative probability to be in a state $s$ where the optimal action is $a$, the algorithm acts to minimize the entropy $H(\phi)$.

Yet another variant, called Entrotaxis, uses the entropy on "predictive measurement distribution as opposed to the entropy of the expected posterior"~\cite{hutchinson2018entrotaxis}. A similar approach to SAI, albeit with quite a different formulation comes from \cite{masson2013olfactory}, which modifies the inference process to reduce the requirements for ``space-processing capabilities'' and proposes a ``mapless search scheme''. The searcher keeps in memory  only the a finite number $M$ of detections alongside their previous position, which are used to construct a posteriori probabilities of the source location via Gaussian mixtures. The searcher then moves accordingly to a target which combines the entropy of source distribution with an explicit term which favours exploitation.


\subsection{Comparison between heuristics and optimal strategies}

Most of the strategies discussed so far, both derived by purely heuristic considerations and from (approximate) optimization, have a long history in robotics and in the olfactory search literature. 

A general, qualitative comparison can already be drawn by looking at the typical shapes of trajectories during the search processes. The simplest heuristic were hard-wired to produce behavioral patterns like surges and casts, mimicking natural behavior. The choice of these features, handpicked by human intuition, seems to have been validated by the following research. In more complex form, the alternation of casts and surges is still recognizable at all levels of algorithm complexity. Infotaxis, and its variants, produces it, as do simpler heuristic such as Thompson sampling, or QMDP. Moreover, they are clearly present for optimal strategies. 

While this qualitative comparison has always been subject of discussion, a consistent, comparative performance was only recently tackled. \cite{loisy2022}, where the SAI heuristic was proposed, demonstrated its superior performance with respect to ``Vanilla" Infotaxis, a result further confirmed in~\cite{heinonen2023optimal}. In the latter work the resulting strategies from Perseus and several heuristics were tested: while Perseus outperformed all heuristics, the much simpler QMDP and Thompson sampling were relatively efficient at large and small emission rates respectively. Different approaches to solve the POMDPs were also compared in~\cite{loisy2023deep}.

\section{Perspectives}\label{sec:polgrad}

Belief update allows for the complete encoding of the past odor history into the spatial maps $b$. However, it requires the knowledge of the observation model, which is not always accessible. Moreover, the biological plausibility that an animal can perform Bayesian computations is a subject of debate.
Interestingly, it is possible to construct and/or optimize different mappings of past histories into coded memories. Given that the agent has the specific goal of searching a task, it may very well be that the relevant information from the past history can be compressed in a much lower-dimensional space than the belief.

One can imagine to replace the belief $b$ and its update operator $\mathcal{T}^{o,a}$ - otherwise uniquely determined by the Bayes' rule - with a different encoding by an abstract vector $\mu$ and a ``memory-update operator" $\mu' = \mathcal{T}^{a,o}_\omega \mu$. As in the previous case, the goal of the learning will be to produce an optimal policy, which now depends on the instantaneous observation and on the current memory state $\pi(a|\mu, o)$.


It is still possible to exploit Reinforcement Learning techniques in POMDPs on these ``memory''-spaces, even though typically these  involve  the Policy Gradient theorem and not explicitly the solution of the Bellman equation. The optimization problem is now effectively divided into two inextricably joint optimization problems, the $\theta$ parameters of the action-selection $\pi(a|o,\mu; \theta)$ and the $\omega$ parameters for the memory-propagation $\mathcal{T}^{a,o}_\omega$. How to construct and update such memories is still an open problem. Possible approaches were outlined in~\cite{meuleau2013learning} and more recently in ~\cite{tottori2022memory}. Here, we will showcase two different approaches which have been recently applied to the olfactory search problem.






\subsection{Recurrent Neural Networks as memory models}

An approach to construct a memory-based process is to use Recurrent Neural Networks (RNN) to encode the memory transition $\mathcal{T}^{a,o}_\omega$. Within this method $\mu$ is a vector in an abstract $\mathcal{R}^d$ space, where generally $d \ll |S|$.   \cite{singh2023emergent} proposes a method to optimize the movement of an agent in an odor plume transported in different turbulent-like conditions of wind, in a model-free context. Analysing the features of the internal representation $\mu$, they recognize the emergence of diverse behavioral ``modes'' which closely resemble a diagram-like behavior. 
A similar approach was also adopted in robotics, where~\cite{hu2019plume} used a Long/short term memory RNN (LSTM) in combination with a Deterministic Policy Gradient algorithm to optimize the movement of an autonomous underwater vehicle in a deep-sea turbulent environment. 

\subsection{Finite-state controller as minimal memories}

Recently \cite{verano2023olfactory} proposed to encode past observations in a discrete set of ``memories'', using a Finite State Controller (FSC)~\cite{hansen2013solving}. In this method, the space $S$ is ``multiplied'' by the number $M$ of memories. At each time the agent is in a spatial position $s$ and in a ``memory'' state $\mu \in \{1,2,\dots,M\}$. The policy, however, depends only on the current memory and the instantaneous observation. 
The optimization process requires to optimize both the policy $\pi(a|\mu,o;\theta)$ \textit{and} the memory transition $\mu' = \mathcal{T}^{a,o}_\omega \mu$. The resulting policy is clearly sub-optimal with respect to policies which depend on the full belief - after all, the belief space is much larger than a finite, discrete set - but it is shown to be robust and very efficient given its very limited computational burden. The optimization process is achieved using the model-based version of the Policy Gradient theorem but it could be in principle be obtained also in a model-\textit{free} setting, where the gradient is stochastically evaluated by experience only, making it relevant for applications in real turbulent flows.

\section{Conclusions}
Olfactory search is the prototypical example of a POMDP that is very hard to solve. The difficulty arises  from the large size of the unobservable state space, i.e. the set of all possible locations of the odor source, and from the paucity of information cues that are available to the searcher when it is far away from the target.

In this chapter we aimed at offering a principled review of some algorithmic approaches to olfactory search. The scope of the methods here described goes well beyond this specific example and naturally extends to more general search processes, including those discussed in this book.

\end{document}